\documentclass[twocolumn,letterpaper]{IEEEAerospaceCLS}  

\usepackage[switch]{lineno}
\usepackage{amsmath}
\usepackage[]{graphicx}    
\usepackage[colorlinks=true, urlcolor=blue, linkcolor=black, citecolor=black]{hyperref} 
\usepackage[sort, numbers]{natbib}

\newcommand{\ignore}[1]{}  
\begin{document}


\title{Reduction of Multiplexer Amplification Noise in Dawn VIR Spectrometer Data }

\author{%
George D. McDonald\\ 
Portland State University\\
1900 SW 4th Ave, Suite 160\\
Portland, OR 97201\\
georgem@pdx.edu
\and 
Kieran Brown\\ 
Portland State University\\
1900 SW 4th Ave, Suite 160\\
Portland, OR 97201
\and 
Bryn King\\ 
Portland State University\\
1900 SW 4th Ave, Suite 160\\
Portland, OR 97201
\and 
Emmanuel Vu\\ 
Portland State University\\
1900 SW 4th Ave, Suite 160\\
Portland, OR 97201
\and 
David Biun\\ 
Portland State University\\
1900 SW 4th Ave, Suite 160\\
Portland, OR 97201
\and 
John Lipor\\ 
Portland State University\\
1900 SW 4th Ave, Suite 160\\
Portland, OR 97201
\and
Ryan S. Park\\
Jet Propulsion Laboratory\\
4800 Oak Grove Dr.\\
Pasadena, CA 91109\\
}

\maketitle

\thispagestyle{plain}
\pagestyle{plain}

\maketitle

\thispagestyle{plain}
\pagestyle{plain}

\begin{abstract}
NASA’s Dawn mission studied the largest main belt asteroids Ceres and Vesta, which inform us of early solar system formation and evolution. Dawn’s Visual and Infrared Spectrometer (VIR) was the primary instrument investigating the mineralogy of these bodies. The spectra taken by VIR have revealed rich deposits of hydrated minerals highly variable at regional scales. However, the VIR spectra contain multiple types of noise which must be corrected prior to analysis. These include odd-even channel noise, spectral spikes, vertical stripes, and artifacts. Odd-even channel noise is caused by the incoming signal being read by IR sensitive cells on two multiplexers, with one reading the evens and another reading the odds. Non-identical material properties of the multiplexers manifest as different gains, with the combined signal appearing to have a saw-tooth pattern. Spectral spikes are due to random events, vertical stripes are due to differences in the relative response of the sample detectors, and artifacts are due to the non-homogeneous instrumental response of the detector. We are carrying out an improved artifact correction procedure to enable machine learning analyses and wider use of the VIR spectra by the planetary science community. We will demonstrate efficacy of applying an artifact correction matrix derived from a statistical sample of observations, specifically in correcting spectral spikes, vertical stripes, and artifacts. We characterize here the performance of this existing method on addressing the odd-even channel noise, and discuss improvements to address this source of noise. Dual-multiplexer spectrometers are commonly flown on spacecraft, and improvements in corrections to the associated odd-even noise has implications for interpretability of the infrared spectra taken by other spacecraft.
\end{abstract} 

\tableofcontents

\section{Introduction}

NASA’s \textit{Dawn} mission studied the largest main belt asteroids Vesta and Ceres, with orbital insertion at Vesta in July of 2011, and the mission ending in Ceres orbit in November of 2018. Dawn’s Visual and Infrared Spectrometer (VIR) was the primary instrument investigating the mineralogy of these bodies. The variety and extent of hydrated minerals on Ceres revealed by VIR are suggestive of recent hydrothermal activity. This has driven increasing interest in Ceres as a potentially habitable body, in turn motivating the recommendation of a NASA Ceres Sample Return in the National Academies' 2023 - 2032 Planetary Science Decadal Survey.

VIR is a dual channel, diffraction grating spectrometer which generates images through pushbroom scanning. The VIS channel operates from 0.25 -- 1.05 $\mu m$, while the IR channel operates in the range of 1.0 -- 5.0 $\mu m$. VIR uses a Shafer telescope and an Offner spectrometer. The detector for the IR channel is a 270 × 435 pixel Mercury Cadmium Telluride array. The first dimension (\textit{samples}) correspond to the across-track spatial dimension. The second (\textit{bands}) are the spectral dimension spanning 432 wavelengths between 1.0 and 5.0 $\mu m$ \cite{desanctis2011}.
The detector is coupled to two multiplexers which respectively read the odd and even columns corresponding to alternate wavelengths \cite{carrozzo2016}. A final dimension, \textit{lines}, consist of separate scans of the pushbroom detector across the surface. Samples, bands and lines serve as the three dimensions for the spectral cube format in which VIR spectra are stored.

Data from VIR has enabled numerous analyses and revolutionized our understanding of Ceres's hydrological history \cite{desanctis2016,raponi2019a,desanctis2020,bramble2022,desanctis2024}. Nevertheless, the VIR data are affected by five types of artifacts. These artifacts are i) an odd-even band (i.e. wavelength) effect in the spectra, essentially a sawtooth pattern ii) spectral spikes, which are spurious strong enhancements or reductions in the signal corresponding to one or a small number of pixels iii) spatial striping i.e. relative differences in the sensitivities of the samples iv) spectral striping i.e. relative differences in the sensitivities of the bands, separate from the odd-even effect and v) a loss of lintearity at the spectrometer's grating filter junctions.

\begin{figure*}
\centering
\includegraphics[width=0.75\linewidth]{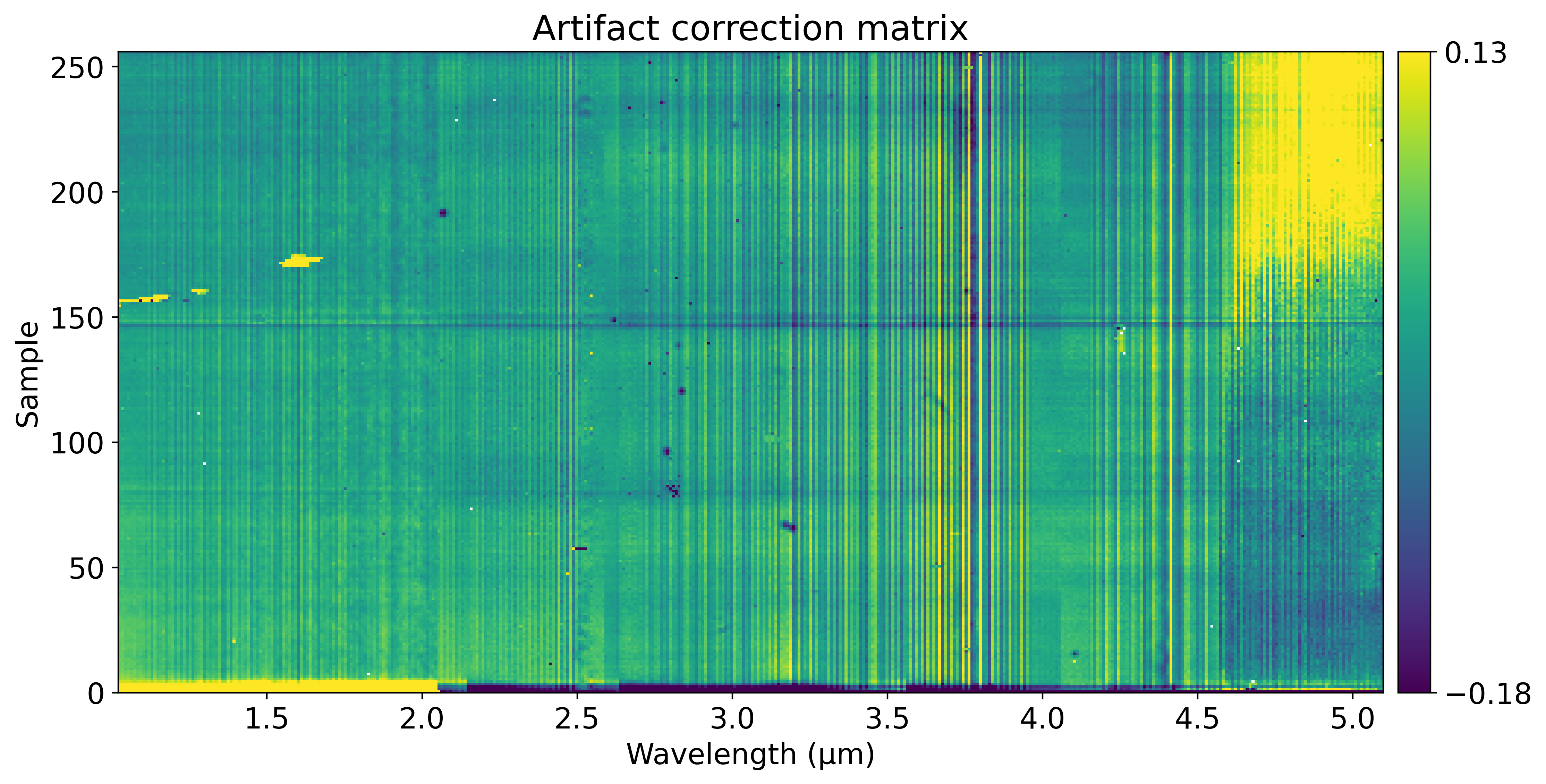}
\caption{The artifact correction matrix for the VIR-IR channel in units of radiance (W/(m$^2$sr$\mu m$))}
\label{fig1}
\end{figure*}

The publicly released VIR spectra in NASA's Planetary Data System (PDS) are calibrated i.e. converted from detector counts to units of radiance (W/(m$^2$ sr $\mu m$)), a measure of the incoming light. Nevertheless the spectra are not corrected for the five artifact types, which affect the calibrated spectra enough to preclude their off-the-shelf use for analyses of surface compositions. To date, scientific interpretation of the VIR data, particularly the IR channel, has been restricted to the VIR team after application of an internal artifact correction procedure. In this work, we implement an improved artifact correction procedure, with the goal of enabling machine learning analyses and wider use of the VIR spectra by the planetary science community. This effort begins with independent construction of an artifact correction matrix, largely following procedures described by the VIR team \cite{carrozzo2016}, and then moves to developing a mathematical model for the pervasive odd-even channel noise (which still manifests, although reduced, after application of the artifact correction matrix) as a function of detector photon counts. The goal is to further reduce the odd-even channel noise.

We focus our correction efforts on VIR's IR Channel, which detects the majority of hydrated minerals observed on Ceres. The \textit{Dawn} at Ceres mission had 5 primary science phases (Approach, RC3, Survey, High Altitude Mapping Orbit [HAMO] and Low Altitude Mapping Orbit [LAM]), and 7 extended mission phases (XM01 -- XM07) \cite{polanskey2016}. We use VIR-IR data from the Survey phase to generate the artifact correction matrix, and generate our photon count odd-even channel noise model using data from the LAMO phase. LAMO is the mission phase with the highest spatial resolution of global regional data usable for scientific analysis, and we test the efficacy of these corrections on LAMO data.

The remainder of this paper is organized as follows.
Section~\ref{artifact} describes the creation of and application of an artifact correction matrix to the VIR-IR data and its effect on the odd-even channel noise.
Section~\ref{photon_count} develops and applies a model for the odd-even channel noise as a function of detector photon count.
Section~\ref{conc} concludes the paper with limitations, improvements being developed, and future work.

\section{Artifact Correction Matrix}
\label{artifact}

The first method we employ to correct the VIR spectra for artifacts is the creation and application of an artifact correction matrix, closely following the methodology described by the VIR team for their internal correction \cite{carrozzo2016}. This matrix is derived from a statistical set of $\sim$20,000 calibrated spectra during Dawn's Survey mission phase, which we refer to as the ``correction dataset''. This procedure, while having demonstrable effects on the odd-even channel noise, addresses four of the five types of noise in the VIR spectra.

\subsection{Creation of the Matrix}

Our interest is in isolating the average manifestation of artifacts. For this reason it is necessary to select a sample of spectra that are i) acquired with relatively low spatial resolution footprints to minimize spatial heterogeneity and ii) lack strong mineral absorptions. Using the Survey mission phase data accomplishes (i), while (ii) requires additional considerations. Specifically, this entails consideration of the spectral characteristics of the target body  at specific wavelength ranges. In this case, observations of Ceres are used from 1.02 -- 2.59 and 4.07 -- 5.1 $\mu m$ where Ceres has few strong absorptions, and for the 2.59 -- 4.07 $\mu m$ range in which Ceres has strong absorptions from hydrated minerals, observations of Vesta are used instead.

The basic procedure is described here only briefly as it is described in detail by Carrozzo et al. \cite{carrozzo2016}. 
\begin{enumerate}
    \item  A median spectrum across the entire correction dataset ($S_{med}(s, \lambda)$) is generated for each sample ($s$) and wavelength ($\lambda$).
    \item The odd-even channel noise is reduced for each $S_{med}(s, \lambda)$ through application of a running three-point weighted average.
    \item Spectral spikes 3$\sigma$ greater than the average signal are replaced with 2nd degree polynomials.
    \item The median of $S_{med}(s, \lambda)$ is taken over the samples, generating unique median spectrum $U_{med}(\lambda)$.
    \item A polynomial function fit ($P_U(\lambda)$) to  $U_{med}(\lambda)$ is calculated.
\end{enumerate}

The artifact correction matrix is then defined as:

\begin{equation}
\label{eq1}
    A(s,\lambda) = \frac{S_{med}(s, \lambda) - P_U(\lambda)}{P_U(\lambda)}
\end{equation}

which is representative of the residual that each sample's average spectrum has with respect to the average across all samples.

\subsection{Application of the Matrix}

Again, following the procedure of Carrozzo et al. \cite{carrozzo2016}:

\begin{enumerate}
    \item Step 2 from the ``Creation of the Matrix'' procedure to reduce odd-even channel noise is repeated, but on the individual spectrum to be corrected.
    \item A new VIR spectrum is calculated (at the individual observation basis, where $l$ is for the line dimension):
    \begin{equation}
        R'(s,l,\lambda) = \frac{R(s,,l,\lambda)}{1 + A(s,\lambda)}
    \end{equation}
\end{enumerate}

\subsection{Characteristics of the Artifact Correction Matrix}

The generated artifact correction matrix $A(s,\lambda)$ is shown in Figure \ref{fig1}. The units, per Equation \ref{eq1}, are in radiance and represent the residual that each sample's average spectrum has with respect to the average across all samples. The dominant residuals are seen to be seen in the wavelength domain, both in terms of large spectral stripes (systematic residuals across multiple wavelengths) and residual odd-even channel effects at the finer scale. As a secondary effect, spatial stripes can be seen in the sample domain. Visually, our artifact matrix compares favorably with the one derived by the VIR team in Figure 7b of Carrozzo et al. \cite{carrozzo2016}.

\subsection{Utility in Correcting Odd-Even Channel Noise}

\begin{figure*}
\centering
\includegraphics[width=\linewidth]{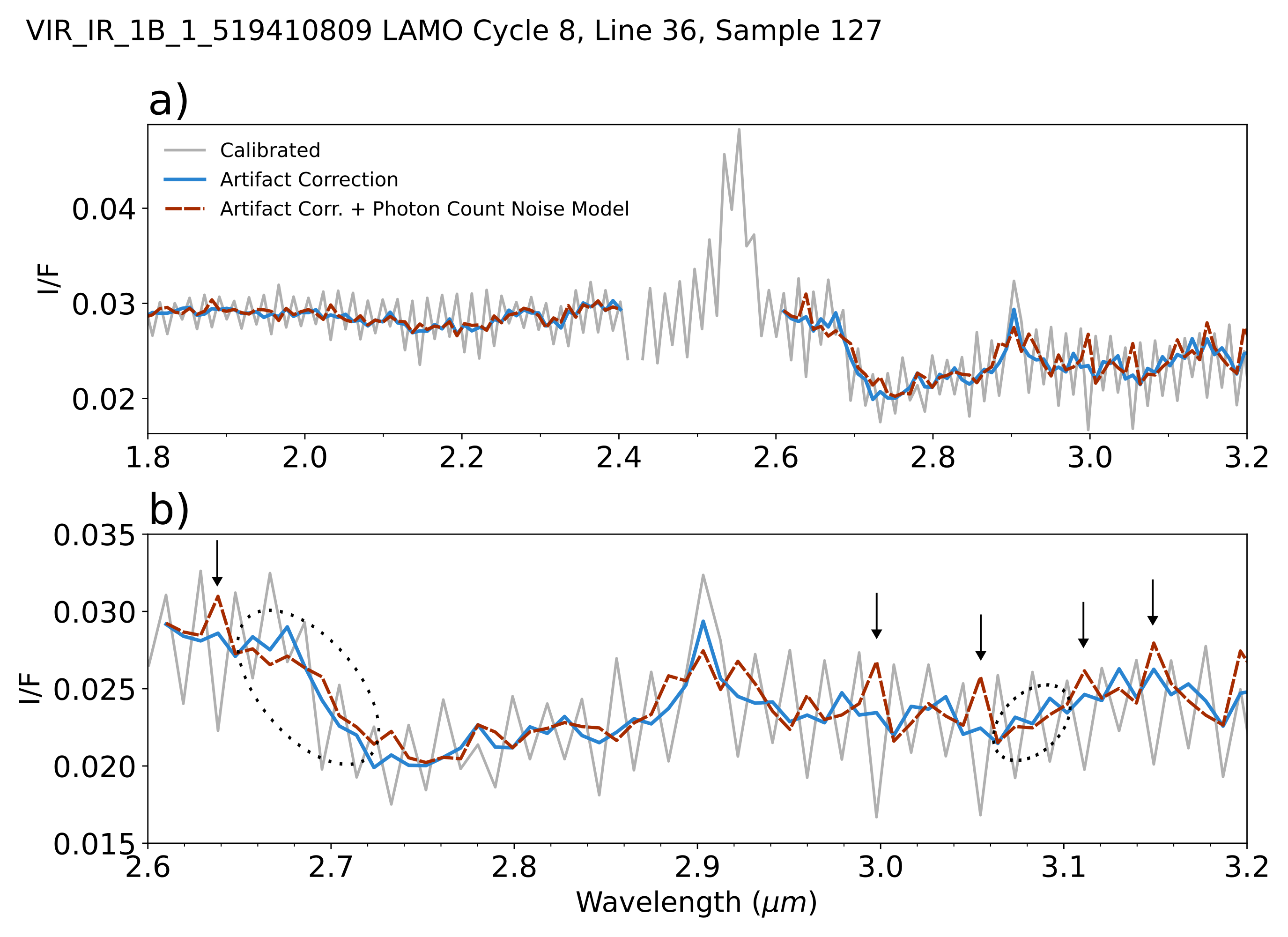}
\caption{\textbf{a)} VIR reflectance spectrum for line 36 and sample 127 of cube LAMO cube 519410809. The original, calibrated radiance spectrum as acquired from NASA's PDS is shown in grey. The blue solid line is the same spectrum after application of the artifact correction matrix. The brown, dashed line is the same spectrum after application of the artifact correction matrix as well as further subtraction of odd-even noise channel noise using the asymmetric Gaussian model \textbf{b)} The same calibrated and corrected spectra, but zoomed in on the 2.6 -- 3.2 $\mu m$ region where hydrated mineral absorptions are present on Ceres. Dashed circles denote regions where the photon count noise model reduces odd-even channel noise compared with simple application of the artifact correction matrix, while arrows denote regions where noise is amplified indicating refinements needed for the model.}
\label{fig2}
\end{figure*}

The result of applying the artifact correction matrix to a single spectrum of Ceres from the Low Altitude Mapping Orbit (LAMO) mission phase can be seen in Figure \ref{fig2}. The spectra here are shown in unitless, relative reflectance (I/F), i.e. the radiance spectrum is divided by the solar radiance spectrum at Ceres, as is done when diagnosing absorptions from surface minerals. The original, calibrated spectrum obtained from the NASA PDS is in light grey, while the spectrum after application of the artifact correction matrix is shown in blue. Panel a) spans the majority of the wavelengths of the IR channel which are unaffected by surface thermal emission, while panel b) focuses on the 2.6 -- 3.2 $\mu m$ region which is used in the analysis of hydrated minerals, particularly hydrated phyllosilicates.

Several large spectral spikes at 3.0 and 3.5 $\mu m$ are removed by the artifact correction matrix. The largest spike at 2.9 $\mu m$ is still present after correction. This is likely a spike which was not affecting the instrument during the Survey phase used to generate the artifact correction matrix, and manifested later in the mission. Otherwise, a factor of four reduction in the odd-even channel noise is seen, from a magnitude of $\sim$0.006 I/F in the original calibrated data to $\sim$0.0015 in the artifact corrected spectrum. Note that the 2.41 to 2.61 $\mu m$ region in the original spectrum is unrecoverable due to a loss of signal linearity at the locations of the junctions of the spectrometer's grating filters \cite{carrozzo2016}.

\section{Photon Count Noise Model}
\label{photon_count}

\subsection{Quantifying the Odd-Even Channel Noise}

The chief, novel contribution of this work is to construct a mathematical model of the manifestation of the odd-even channel noise, which can then be used to mitigate it effects. Part of the difficulty in correcting this noise is that its magnitude is variable over time and amongst individual observations. In order to derive such a model, we first investigated physical variables with which the noise might be correlated, namely the temperature of the detector, and the incoming photon flux.

To examine these correlations, we quantify the magnitude of the odd-even channel noise. Conceptually, this is the magnitude of the of the odd-even perturbations (in radiance) with respect to the underlying overall spectrum. This is done using a same-parity residual filter. For each wavelength channel $\lambda$, an estimate of the baseline spectrum is constructed by averaging the second-nearest neighbors ($\lambda$ - 2 and $\lambda$ + 2), to preserve the parity between odd and even channels. Subtraction of this same-parity baseline from the overall spectrum isolates high-frequency channel-to-channel variations ($X_\lambda = S_\lambda - \hat{S}_\lambda$) as a local noise residual. In summary, this is:

\begin{equation}
    X_\lambda = S_\lambda - \hat{S}_\lambda = S_\lambda - \frac{S_{\lambda-2} + S_{\lambda+2}}{2}
\end{equation}

With temperature, we find no meaningful correlation with this noise $X\_\lambda$. Because the detector temperatures reported in the PDS are quantized in increments of 0.5 K, partial correlations with small changes in temperature cannot be ruled out---nevertheless there is not a discernible trend in $X_\lambda(T)$ is.

\subsection{Correlation of the Odd-Even Channel Noise with Photon Count}

We next examined correlation of the odd-even channel noise $X_\lambda$ with detector photon counts. Detector photon counts were calculated from radiance in the calibrated spectra using the etendue of VIR, as well as the exposure time (and the $hc$ constant, i.e. wavelength-energy conversion factor). Because VIR employs a dual-amplifier design, where odd and even spectral channels of the detector are passed through distinct readout amplifiers, there is physical motivation for $X_\lambda$ also increasing with the magnitude of the incoming signal \cite{desanctis2011, carrozzo2016}. When plotting the entire set of observations by a single detector pixel (s, $\lambda$) over the LAMO mission phase, this growth in the odd-even channel noise with photon count is apparent (the grey points in Figure \ref{fig3}a and b). The form of this noise can be interpreted as follows: i) a \textit{baseline floor} representing an intrinsic static offset residual during low-signal ii) a \textit{ramp} region where as the photon count $q$ increases toward saturation, the odd-even channel noise grows exponentially or close to exponentially and iii) a \textit{peak artifact amplitude} marking a critical photon count where the amplifier path hits full-well saturation. While not apparent for the pixels in Figure 3, certain pixels exhibit iv) a \textit{post-saturation collapse} which is inferred to be a charge overflow from the pixel of interest into adjacent pixels during the readout phase. This follows descriptions of similar odd-even noise in other mercury cadmium telluride detectors for infrared spectrometers \cite{yanagisawa2006conf}.

\subsection{Asymmetric Gaussian Model for the Odd-Even Channel Noise}

Because we aim to model $X_\lambda(q)$ so as to best isolate its effects, we require a mathematical function that can describe these four characteristics of the odd-even channel noise. We employ an asymmetric Gaussian function, as its bell-shaped profile is able to capture variable saturation thresholds, provide an exponential transition between the baseline floor and peak artifact amplitude, and also allow for a post-saturation collapse with a different slope. Mathematically, this is:

\begin{equation}
X_i(q) = A + C \cdot \exp\left( -\frac{1}{2} \left[ \frac{q - q_{\text{peak}}}{\sigma(q)} \right]^2 \right)
\end{equation}

\begin{equation}
\sigma(q) = 
\begin{cases} 
\sigma_l, & \text{if } q < q_{\text{peak}} \\ 
\sigma_r, & \text{if } q \ge q_{\text{peak}} 
\end{cases}
\end{equation}

where $A$ represents the baseline floor noise value, $C$ is the peak artifact amplitude at photon count $q_{\text{peak}}$, $\sigma_l$ is the slope for the ramp region, and $\sigma_r$ is the slope for the post-saturation collapse. With this function, we are able to model a representative $X_i(q)$ for each pixel ($\lambda$, $s$) over the LAMO mission phase. $X_i(q)$ can then be evaluated for the photon count of an individual observation to be corrected, and then this noise can be subtracted from the overall signal in radiance. We note that the Gaussian function is a phenomenological model, as the true microscopic physics of charge transfer, sense node capacitance saturation, and potential A/D rail clipping involve step-like thresholds and non-linear semiconductor equations. However, the expectation (and observation) is that in ensemble behavior across hundreds of thousands of readouts, the noise manifests as a smooth, bell-like increase and collapse with photon count. Such an asymmetric Gaussian model has previously been applied to detector response in X-ray spectrometers \cite{roseker2011}.

\subsection{Applying the Photon Count Noise Model to Spectra}

The impact of subtracting the asymmetric Gaussian modeled odd-even channel noise, as applied to a single spectrum (and after prior application of the artifact correction matrix) is shown in Figure \ref{fig2} as the brown, dashed line. The dashed circles denote spectral regions where a noticeable decrease in odd-even channel noise is observed. In contrast, the arrows indicate regions where the odd-even channel noise has been amplified by application of the asymmetric Gaussian model. The fact that improvements are observed in certain spectral regions suggests that this model does have potential for mitigating the odd-even channel noise. Nevertheless, the introduction of artifacts in places indicates that refinements to the model for $X_\lambda(q)$, and its application are needed.

\begin{figure}\label{fig3}
\centering
\includegraphics[width=3.25in]{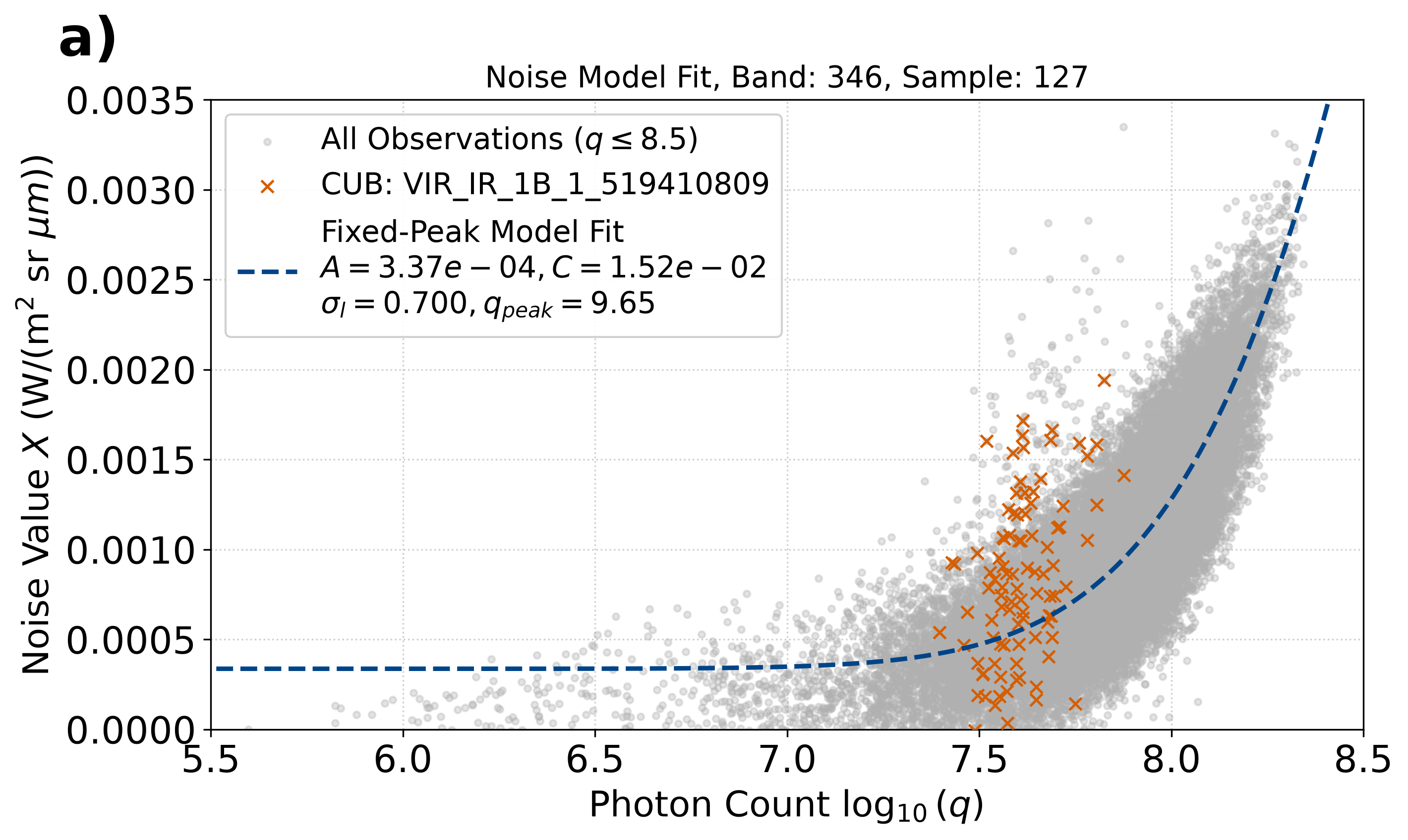}\\
\includegraphics[width=3.25in]{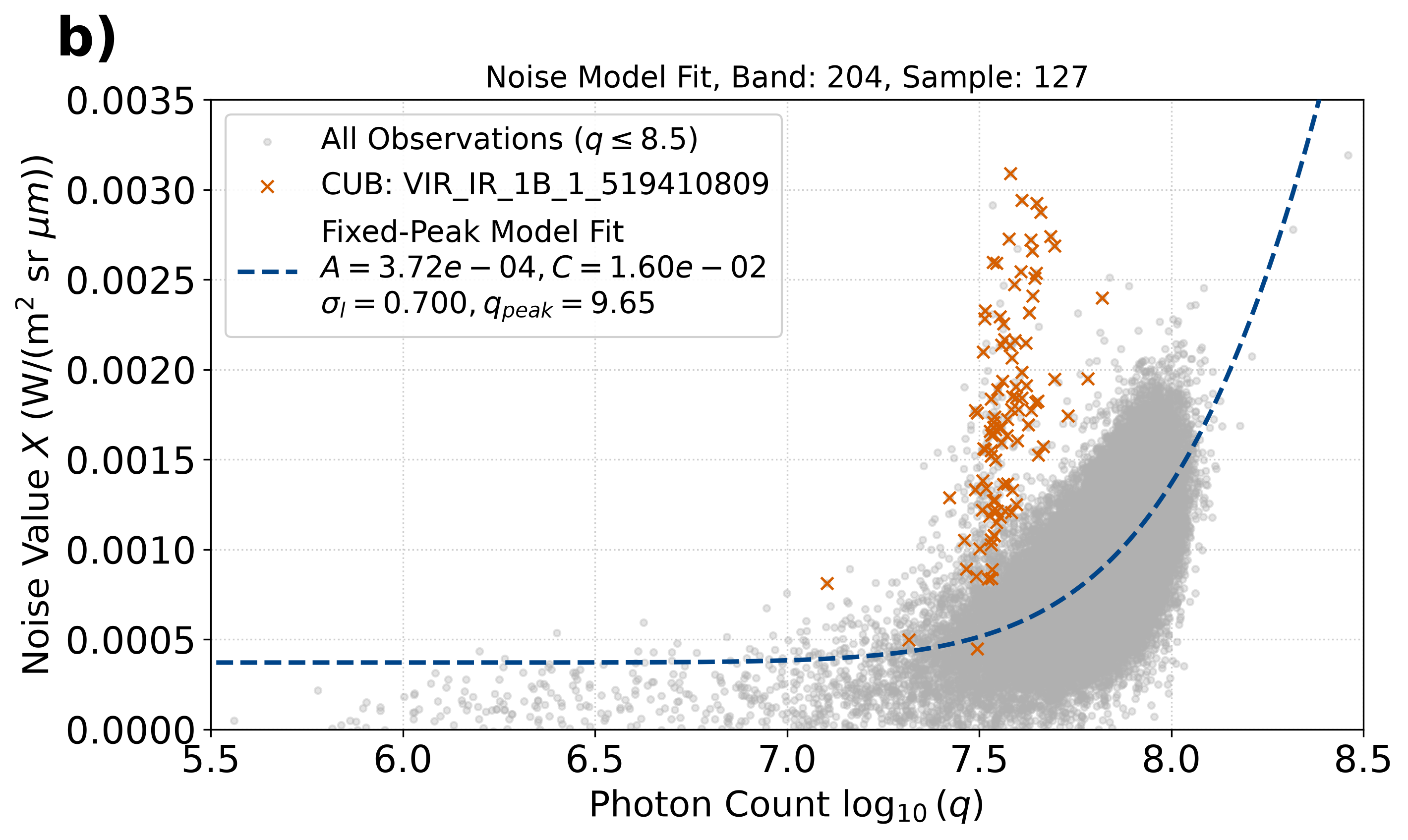}\\
\caption{Individual measurements of odd-even channel noise for as a function of the logarithm of the photon count, for two individual detector pixels over the entirety of LAMO observations. The asymmetric Gaussian fit to the observations is shown as the blue dashed lines. The noise vs photon count for VIR observation 519410809 of Figure 2 is shown as the orange X's.}
\end{figure}

We have identified two effects that can result in the subtraction of $X_\lambda$, as currently modeled, actually enhancing odd-even channel noise. Firstly, it can be seen that all of the arrows in Figure \ref{fig2} point to regions where the original magnitude of the odd-even channel noise is negative with respect to the baseline of the spectrum. The asymmetric Gaussian model, as currently employed, demonstrates difficulties in accurately fitting peaks with negative amplitudes (presumed to be an issue with the non-linear least squares optimization routine). We are examining improvements to the fitting procedure that are expected to enable accurate modeling of negative peaks---the fact that these introduced artifacts are unique to negative signs of the odd-even channel noise is promising for prospects of their elimination in future versions of the algorithm. A second effect, is a result of the assumption made when fitting $X_\lambda$ to the entire set of LAMO spectra for a given sample and band, that the spectrum being corrected exhibits odd-even channel noise that does not deviate substantially from the average behavior. $X_{\lambda}$(q) for the VIR observation of Figure \ref{fig2} (spectral cube 519410809), is plotted as orange X's, separately from the overall LAMO dataset for two pixels in Figure \ref{fig3}. While at spectral band 347 and sample 127 (Figure \ref{fig3}a), the noise trends for this cube are a subset of the overall trends for LAMO, at spectral band 204 and sample 127 (Figure \ref{fig3}b), it is apparent that $X_{\lambda}$(q) differs for this cube from the overall trend in the LAMO mission phase. For such a scenario, an attempt to correct $X_{\lambda}$(q) using the average LAMO behavior will fail to accurately model the noise for this individual observation. Finally, it can be seen that there is significant scatter in $X_{\lambda}$(q), such that the efficacy of the correction of the odd-even channel noise modeled as a function of photon counts will always have intrinsic variability.

\section{Conclusions}
\label{conc}

We have carried out an independent effort to correct the \textit{Dawn} spacecraft's Visual and Infrared Spectrometer (VIR) IR channel data for artifacts. The generation and successful application of an artifact matrix, while largely following procedures described by the VIR team \cite{carrozzo2016}, successfully mitigates four out of the five types of artifacts present in the VIR-IR data, enabling for the first time scientific analysis of the VIR-IR data outside of the VIR team.

Our metric for quantification of the magnitude of VIR's odd-even channel noise ($X_\lambda$), and investigation of its functional form as dependent on detector photon counts is novel. An asymmetric Gaussian function shows promise in encapsulating the ensemble behavior of $X_\lambda$. While the modeling out of the odd-even channel noise using this function is currently mixed in its efficacy, this has been tied to the challenges of fitting noise with both positive and negative magnitudes, and achieving good, unique fits across all 432 spectral bands. Improvements to our algorithm as related to this simultaneous fitting across bands are expected to produce a model which consistently reduces the odd-even channel noise for a majority of wavelengths.

One challenge of the odd-even noise is that even with an accurate and well-performing photon count ($q$) model, there is significant scatter in $X_\lambda(q)$, indicating that although measurable reductions in its manifestation are possible, this method will not eliminate it. We are using additional observations, namely VIR in-flight calibration images of a polystyrene filter, as well as images of standard stellar sources, to examine additional variables that may explain part of the scatter in the odd-even channel noise as a function of photon count.

Our artifact-corrected VIR spectra are already being used for novel spectral analyses using machine learning \cite{biun2027}, and we intend to release the final versions of the artifact-corrected VIR-IR spectra to the wider planetary science community with a future scientific publication.

\section{Acknowledgements}

This work was supported by NASA Discovery Data Analysis Program 80NSSC25M0030, and in part through NASA and Oregon Space Grant Consortium, Cooperative Agreement
80NSSC25M7044.
The authors used Gemini 3.6 Flash (Google) to help write the code for asymmetic Gaussian function curve fitting, generating figures, as well as generating LaTeX equations. All AI-generated content was reviewed and edited by the authors, who take full responsibility for the content of this paper.

\section{References}
\bibliographystyle{IEEEtran}
\bibliography{library_26}

\thebiography

\begin{biographywithpic}
{George D. McDonald}{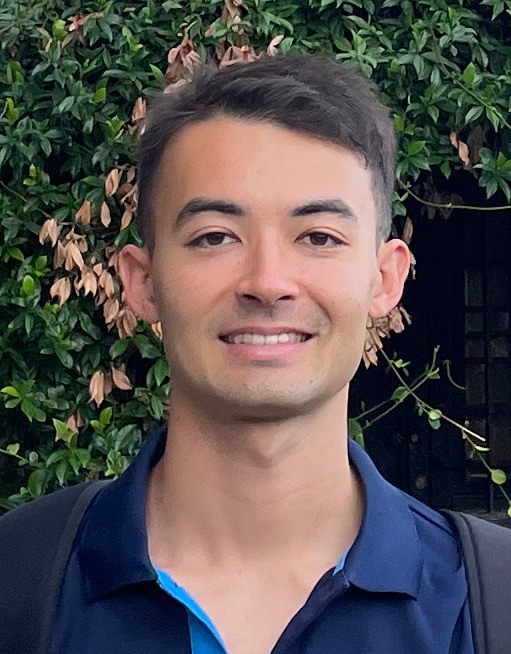} is currently a Research Assistant Professor of Electrical \& Computer Engineering at Portland State University. He is a Collaborator on the Mastcam-Z and Mars Environmental Dynamics Analyzer (MEDA) instruments on NASA's \textit{Perseverance} rover, as well as a Collaborator on the Rover Environmental Monitor Station (REMS) on the \textit{Curiosity} rover. He received a Career Development Award from the Lunar and Planetary Institute in 2015. His primary research uses imagery, spectroscopy, and in-situ sensing to understand modern-day environmental changes on the terrestrial planets, as well as using numerical simulations to understand planetary boundary layers.
\end{biographywithpic}

\begin{biographywithpic}
{Kieran Brown}{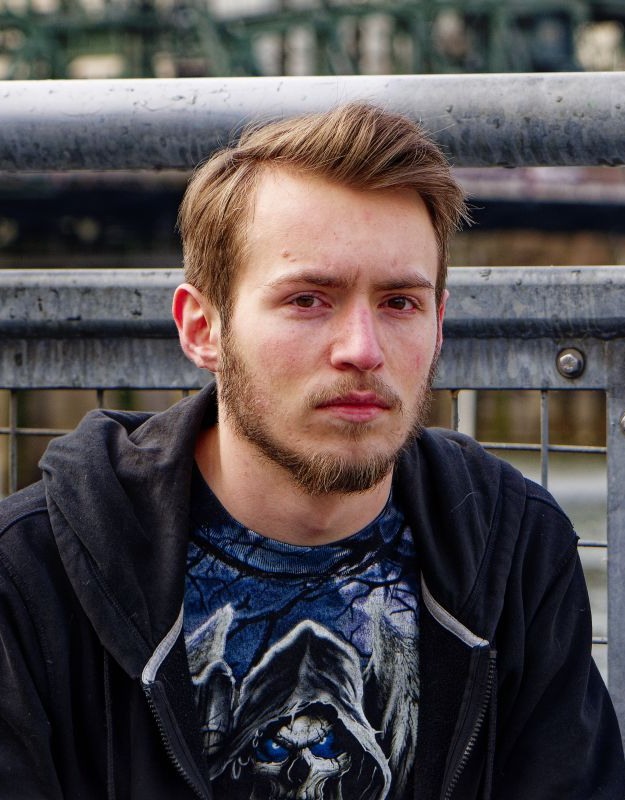}
is an undergraduate student at Portland State University majoring in Computer Engineering.
With a lifelong passion for computers and science, he has an interest in the development and optimization of tooling and processing pipelines for remote sensing work. He is currently a Student Collaborator on the Mastcam-Z camera team on NASA's \textit{Perseverance} rover.
\end{biographywithpic}

\begin{biographywithpic}
{Bryn King}{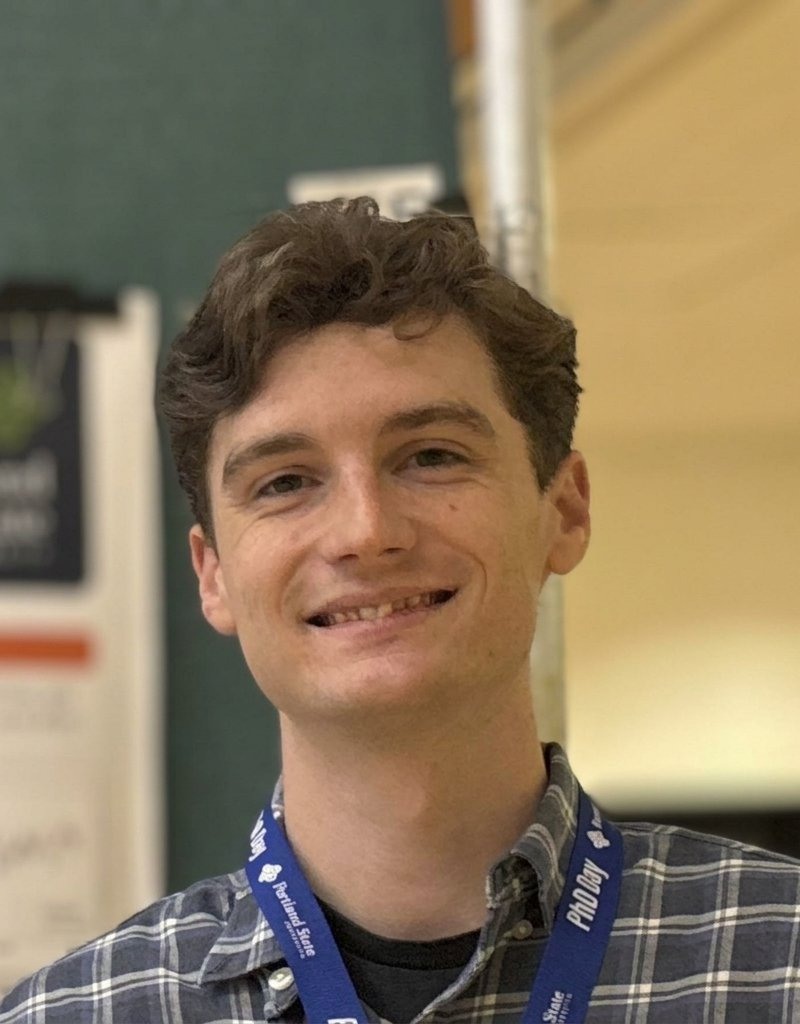}
is an undergraduate research assistant in the Environmental Sensing Lab at Portland State University, with a focus on signal processing and imaging techniques.
He is pursuing his bachelor's degree in Electrical Engineering.
Areas of interest include automation and control systems.
\end{biographywithpic}

\begin{biographywithpic}
{Emmanuel Vu}{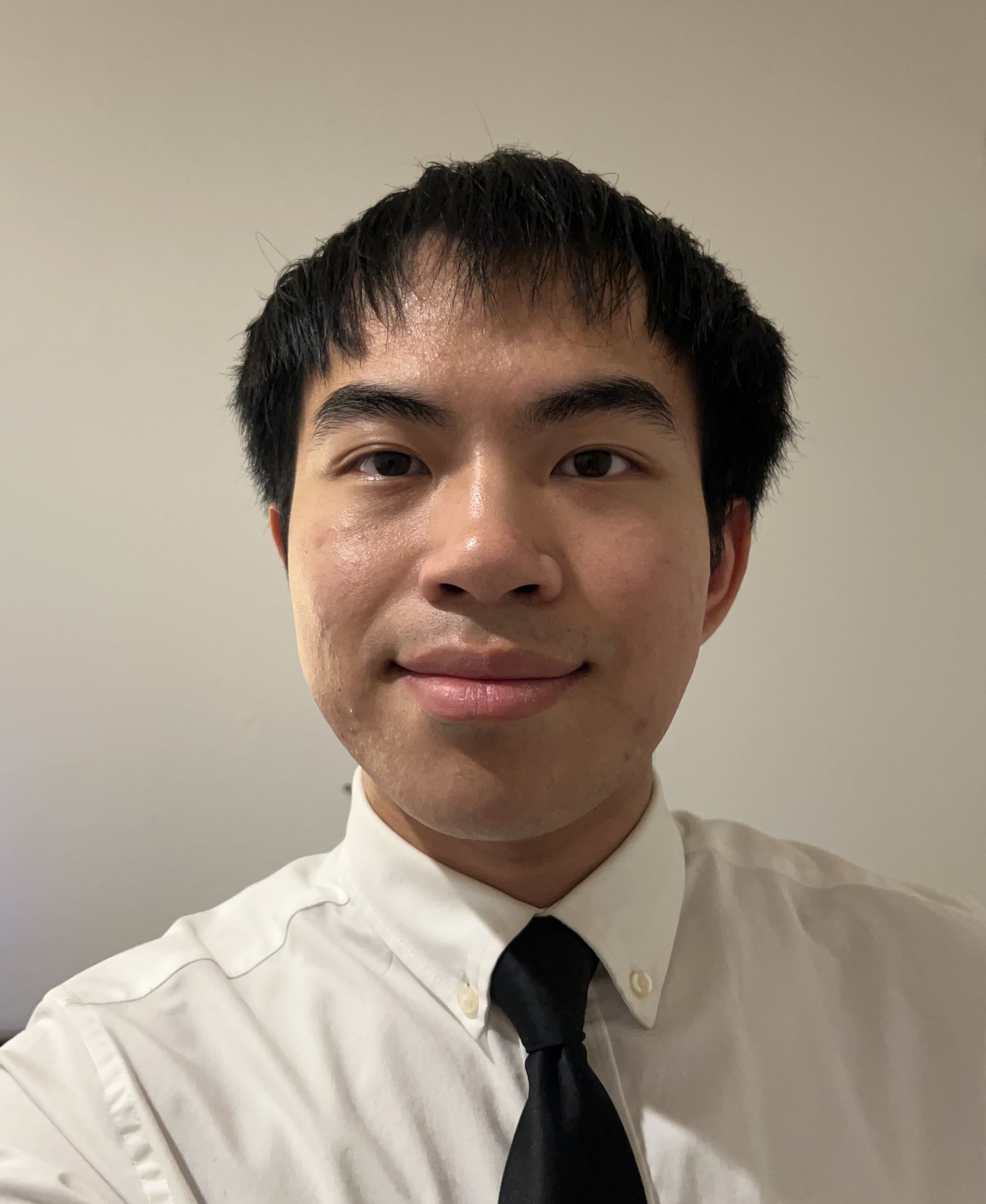}
is currently pursuing a Bachelor of Science in Electrical Engineering at Portland State University. He intends to graduate in the spring of 2026. 
\end{biographywithpic}

\newpage

\begin{biographywithpic}
{David Biun}{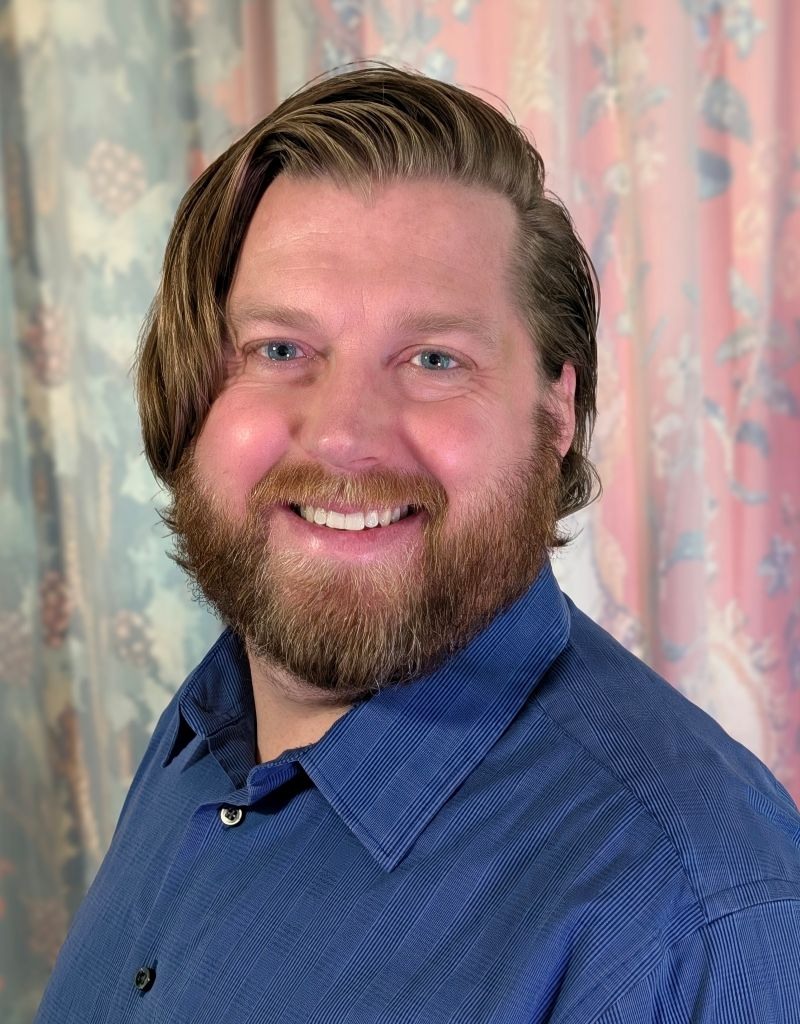}
received his B.A.\ in Music from the University of Rochester before moving to a career in winemaking for over a decade.
He is currently in the final year of his Master's degree in Electrical and Computer Engineering at Portland State University.
His coursework is focused on computer vision and machine learning, and he is preparing to defend his thesis on subspace clustering algorithms.
\end{biographywithpic}

\begin{biographywithpic}
{John Lipor}{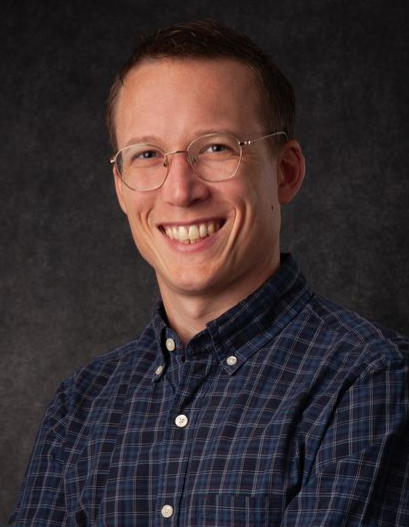}
is an Associate Professor Electrical \& Computer Engineering and a Distinguished
Research Scientist at Metron, Inc. He is a recipient of the Defense Advanced Research
Projects Agency (DARPA) Young Faculty Award, the DARPA Director’s Fellowship, and
the National Science Foundation CAREER Award. Dr. Lipor’s primary research focuses
on active and adaptive methods in machine learning, with an emphasis on applications in
environmental sensing and monitoring.
\end{biographywithpic}

\begin{biographywithpic}
{Ryan S. Park}{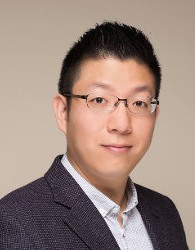}
is a Principal Engineer and Supervisor of the Solar System Dynamics Group at NASA's Jet Propulsion Laboratory.
His group predicts the orbits of planets, planetary moons, asteroids, and comets for space missions and researchers worldwide.
His research focuses on spacecraft navigation, satellite geodesy, and planetary dynamics, with an emphasis on high-precision estimation and dynamical modeling for planetary exploration.
He has contributed to numerous space missions, serving as a Co-Investigator on Juno, Psyche, and Europa Clipper, and as Principal Investigator of the Advanced Pointing Imaging Camera and Gravity Imaging Radio Observer.
Dr.\ Park is an Adjunct Professor at the University of Southern California and a Fellow of the American Astronautical Society.
\end{biographywithpic}

\end{document}